\documentclass{emulateapj}

\usepackage{multirow}
\usepackage{amsmath}
\usepackage{longtable}

\newcommand{\scl}	{\Sigma_{\rm cl}}
\newcommand{\gcm}	{{\rm g}\,{\rm cm}^{-2}}
\newcommand{\cogas}{\mathrm{CO}_\mathrm{gas}}
\newcommand{\tauon}{\tau_\mathrm{on}}
\newcommand{\tauoffcr}{\tau_\mathrm{off,CR}}
\newcommand{\tauoffth}{\tau_\mathrm{off,th}}

\shorttitle{Environment and Protostellar Evolution}
\shortauthors{Zhang \& Tan}

\begin{document}

\title{Environment and Protostellar Evolution}

\author{Yichen Zhang$^{1,2}$ \& Jonathan C. Tan$^{3,4}$}
\affil{$^1$Departamento de Astronom\'ia, Universidad de Chile, Casilla 36-D, Santiago, Chile\\	
$^2$Department of Astronomy, Yale University, New Haven, CT 06520, USA\\
$^3$Department of Astronomy, University of Florida, Gainesville, Florida 32611, USA\\
$^4$Department of Physics, University of Florida, Gainesville, Florida 32611, USA\\
yczhang.astro@gmail.com}

\begin{abstract}
Even today in our Galaxy, stars form from gas cores in a variety of
environments, which may affect the properties of resulting star and
planetary systems. Here we study the role of pressure, parameterized
via ambient clump mass surface density, on protostellar evolution and
appearance, focussing on low-mass, Sun-like stars and considering a
range of conditions from relatively low pressure filaments in Taurus,
to intermediate pressures of cluster-forming clumps like the Orion
Nebula Cluster (ONC), to very high pressures that may be found in the
densest Infrared Dark Clouds (IRDCs) or in the Galactic Center
(GC). We present unified analytic and numerical models for collapse of
prestellar cores, accretion disks, protostellar evolution and bipolar
outflows, coupled to radiative transfer (RT) calculations and a simple
astrochemical model to predict CO gas phase abundances. Prestellar
cores in high pressure environments are smaller and denser and thus
collapse with higher accretion rates and efficiencies, resulting in
higher luminosity protostars with more powerful outflows.  The
protostellar envelope is heated to warmer temperatures, affecting
infrared morphologies (and thus classification) and astrochemical
processes like CO depletion to dust grain ice mantles (and thus CO
morphologies). These results have general implications for star and
planet formation, 
especially via their effect on astrochemical and dust grain evolution
during infall to and through protostellar accretion disks.
\end{abstract}


\section{Introduction}
\label{sec:intro}

In self-gravitating virialized gas clumps, internal pressure is set by
gas envelope weight, which is related to the more easily observable
mass surface density, $\scl$, via $P\simeq\:G\scl^2$
(\citealt[]{MT03}, hereafter MT03). Observed $\Sigma$s of star-forming
regions vary greatly, from regions like Taurus with
$\Sigma\sim0.03\:\gcm$ (\citealt[]{Onishi96}) where low-mass
stars are forming in relative isolation, to regions like the ONC with
$\Sigma\sim0.3\:\gcm$ (MT03) where low-mass stars are more crowded
around massive stars. IRDCs have $\Sigma\sim0.1-1\;\gcm$
(\citealt[]{BT12}). Some massive-star-forming regions reach
$\Sigma\gtrsim1\;\gcm$ (\citealt[]{Plume97,Battersby14,Tan14}).
Here low-mass stars are also expected to form
along with massive stars from the fragmenting clump. GC region clouds,
such as the ``Brick,'' also contain clumps of very high $\Sigma$
(\citealt[]{Rathborne14}). How do these different environmental
conditions affect formation of Sun-like stars?

In the Turbulent Core model for massive star formation (MT03), initial
core size and subsequent protostellar evolution are determined by
initial core mass, $M_c$, and $\Sigma$ of surrounding
clump. This model can be extended to low-mass star formation, assuming
cores are approximately virialized and in pressure balance with the
clump. A prestellar core, i.e., on verge of collapse, of given mass in
a higher $\Sigma$ environment experiences a higher surface pressure,
is therefore more compact and dense, and thus collapses with higher
accretion rate. Higher luminosity due to faster accretion and more
compact structure make the core warmer.
Such dependence of protostellar core temperature on environment has
further implications for infrared morphologies, chemical evolution of
core and disk, and potentially planet formation.

The above scenario falls within the ``core accretion'' paradigm, in
which accretion rate and final stellar mass are determined by initial
conditions of the core. 
In the alternative ``competitive accretion'' paradigm
(\citealt[]{Bonnell01}), there is no such dependence on initial core
properties.

Some observations of protoclusters suggest neighboring protostars
have more correlated luminosities and accretion rates
(\citealt[]{Kryukova14}, \citealt[]{Elmegreen14}).  This has been
explained invoking mass segregation and large-scale accretion flows,
which appear in competitive accretion, but may also result from
accretion rate dependence on star formation environment as predicted
by core accretion. A better quantitative understanding of this
environmental dependence will help solve this question.

Here we study how star-forming environment ($\scl$) affects thermal
evolution, infrared morphologies and CO gas phase abundances of
protostellar cores.
The infrared appearance of low-mass protostars and its dependence on
envelope, disk, and outflow properties were studied by
\citet[]{Robitaille06} with a suite of RT models. Temperature and
chemical evolution of protostellar cores have also been studied using
various astrochemistry models incorporated into either analytical
models of collapse (e.g.,~\citealt[]{Visser09,Visser11}) or dynamical
simulations (e.g.,~\citealt[]{Hincelin13}).  However, dependence of
temperature evolution, infrared morphology, and chemistry on initial
environmental conditions, especially pressure and $\Sigma$, has not
yet been studied with a self-consistent model including collapsing
cores, disks, gradually widening outflow cavities and full
protostellar evolution. This is our goal.

\section{Models}
\label{sec:model}

%========================================
\begin{figure*}
\begin{center}
\includegraphics[width=0.9\textwidth]{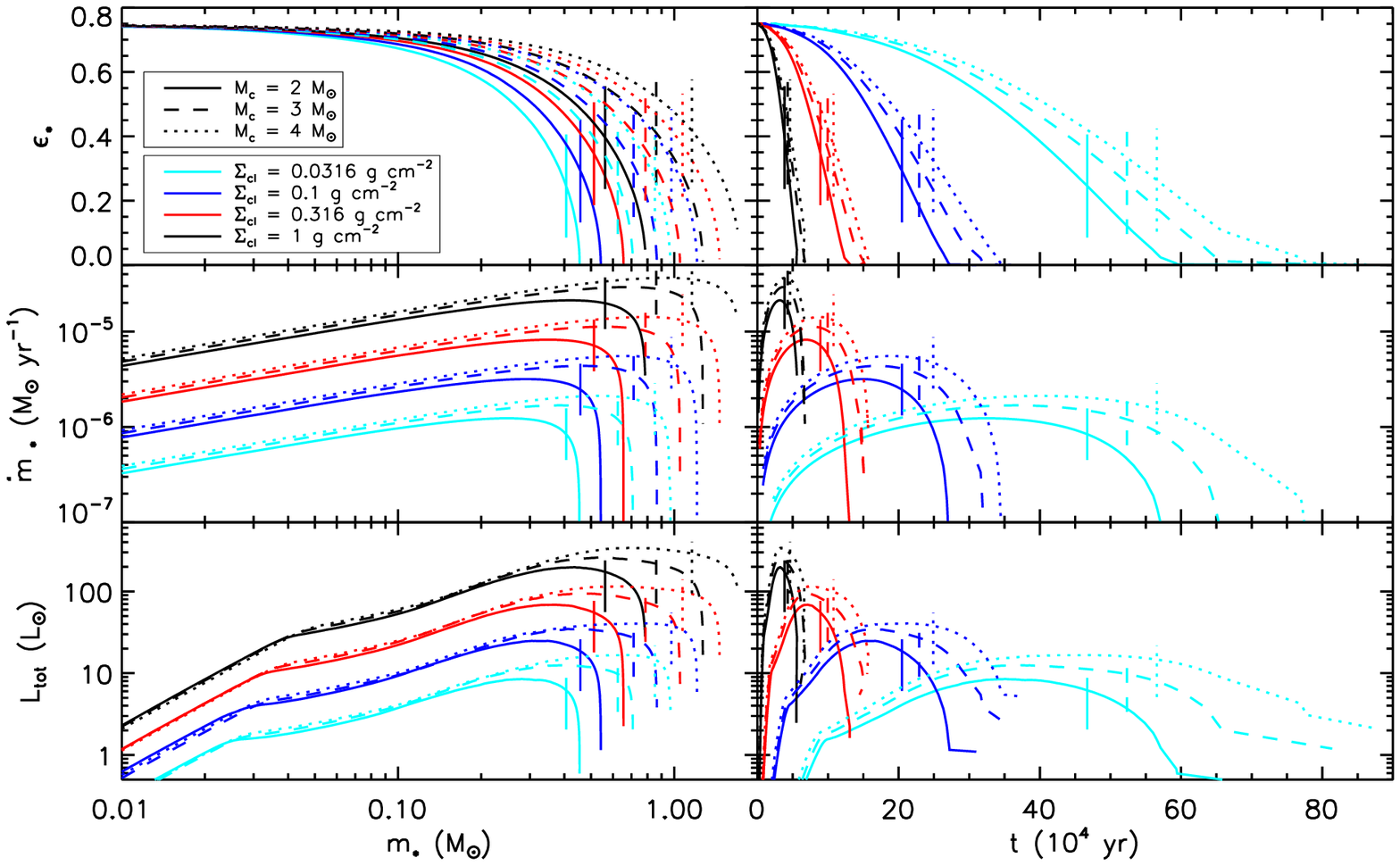}\\
\caption{Evolution of star formation efficiency (top), stellar
accretion rate (middle), and total luminosity (bottom) 
in models with various $\scl$ and $M_c$ (see
legend) versus stellar mass (left column) and time (right column).
The short vertical lines indicate when $m_*=M_{\rm{env}}$.}
\label{fig:history}
\end{center}
\end{figure*}
%========================================

%========================================
\begin{figure*}
\begin{center}
\includegraphics[width=0.9\textwidth]{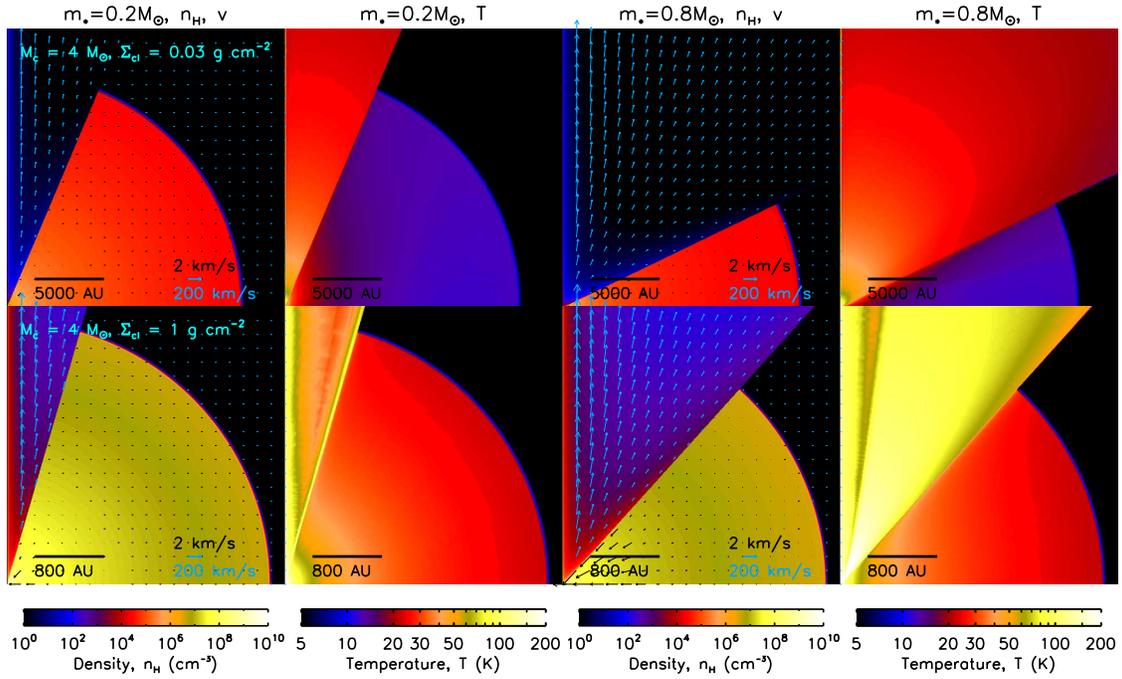}\\
\caption{RT simulation input density profiles (1st/3rd columns) and converged
temperature profiles (2nd/4th columns) for $M_c=4\;M_\odot$ cores at
$m_*=0.2\:M_\odot$ (1st/2nd columns) and $m_*=0.8\:M_\odot$ (3rd/4th
columns) for $\scl=0.03\;\gcm$ (top row) and $\scl=1\;\gcm$ (bottom row).
In each panel, protostar is at lower-left, x-axis follows disk
midplane, and y-axis follows outflow/rotation axis. Note, upper and
lower rows have different size scales. Blue/black arrows show
outflow/inflow velocity fields, with different scales.}
\label{fig:trho3d}
\end{center}
\end{figure*}
%========================================

%========================================
\begin{figure}
\begin{center}
\includegraphics[width=0.95\columnwidth]{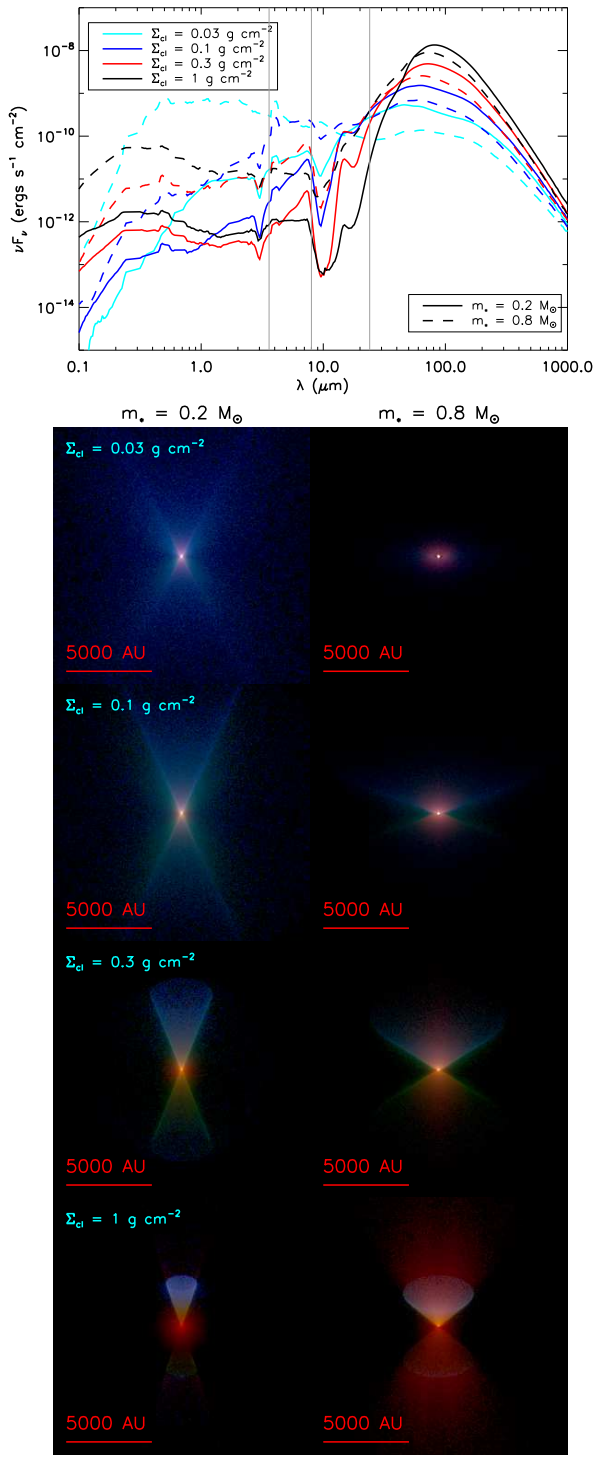}\\
\caption{SEDs and 3-color images (blue/green/red: $3.6/8/24\:\micron$)
for $M_c=4\;M_\odot$ cores viewed at $500\;\mathrm{pc}$ distance and
$60^\circ$ inclination angle between line-of-sight and outflow axis
with $m_*=0.2\:M_\odot$ (left) and $m_*=0.8\:M_\odot$ (right) 
for $\scl=0.03,\:0.1,\:0.3,\:1\;\gcm$ (top to bottom rows).
One color stretch is used for $m_*=0.2\:M_\odot$ models,
another for $m_*=0.8\:M_\odot$ models. So relative color comparison
should be made only for models at the same evolutionary stage.}
\label{fig:sedimg}
\end{center}
\end{figure}
%========================================

%========================================
\begin{figure*}
\begin{center}
\includegraphics[width=0.8\textwidth]{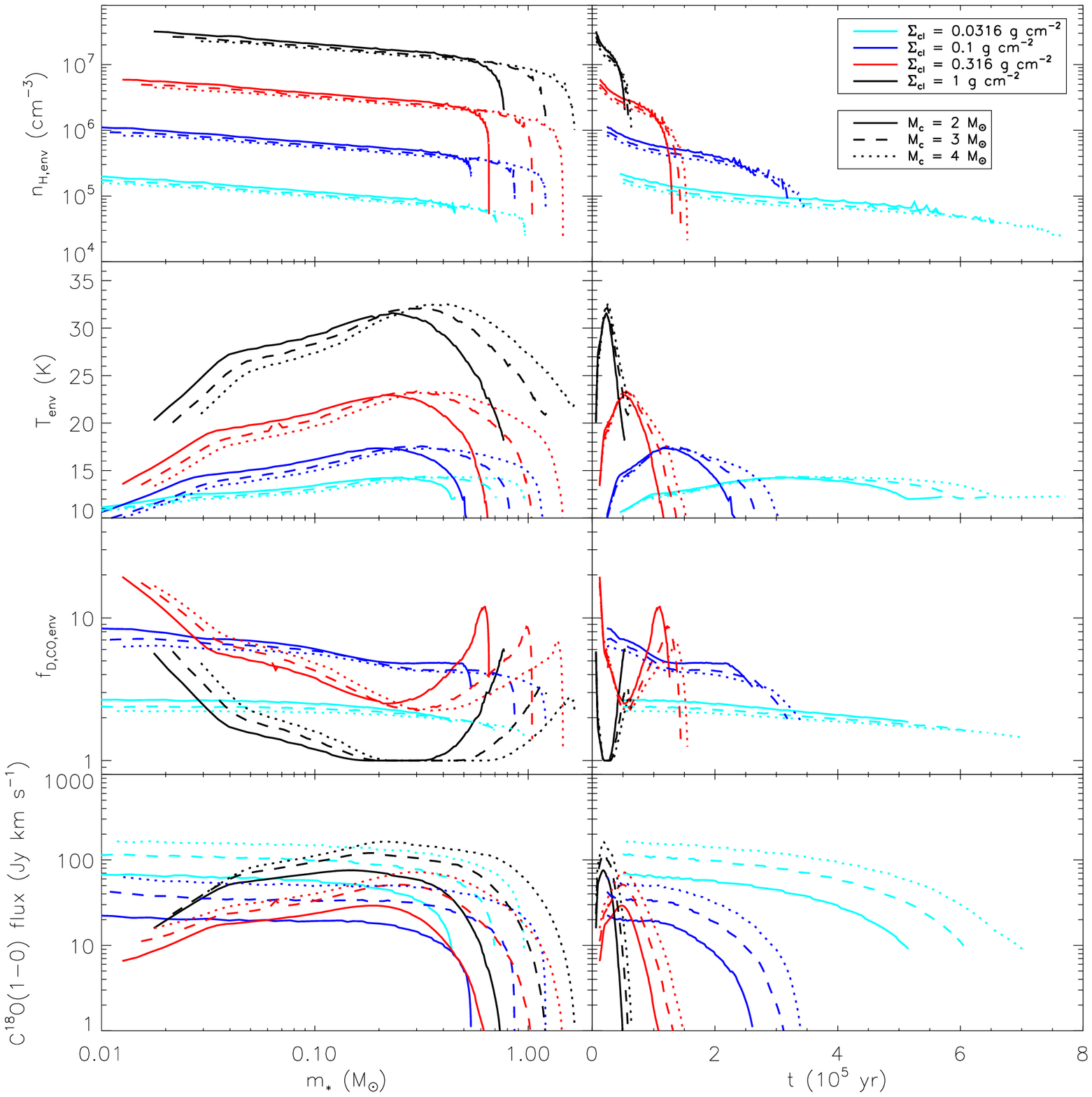}\\
\caption{Evolution of envelope mass-weighted mean density 
(top row), mean temperature ($T_\mathrm{env}$, 2nd row), 
CO depletion factor 
($f_{D,\mathrm{CO,env}}=M_{\mathrm{CO,env}}/M_{\mathrm{CO,gas,env}}$, 3rd
row), and C$^{18}$O($J$=1-0) flux (bottom row) 
with stellar mass (left) and time (right) for models with various $\scl$ and $M_c$.
The C$^{18}$O abundance is calculated from [$^{12}$CO]/[H$_2$]=$10^{-4}$ (\citealt[]{Frerking82}), 
[$^{12}$CO]/[C$^{13}$O]=62 (\citealt[]{Langer93}) and [$^{13}$CO]/[C$^{18}$O]=5.5 (\citealt[]{Wilson92}).}
\label{fig:evo}
\end{center}
\end{figure*}
%========================================

%========================================
\begin{figure}
\begin{center}
\includegraphics[width=\columnwidth]{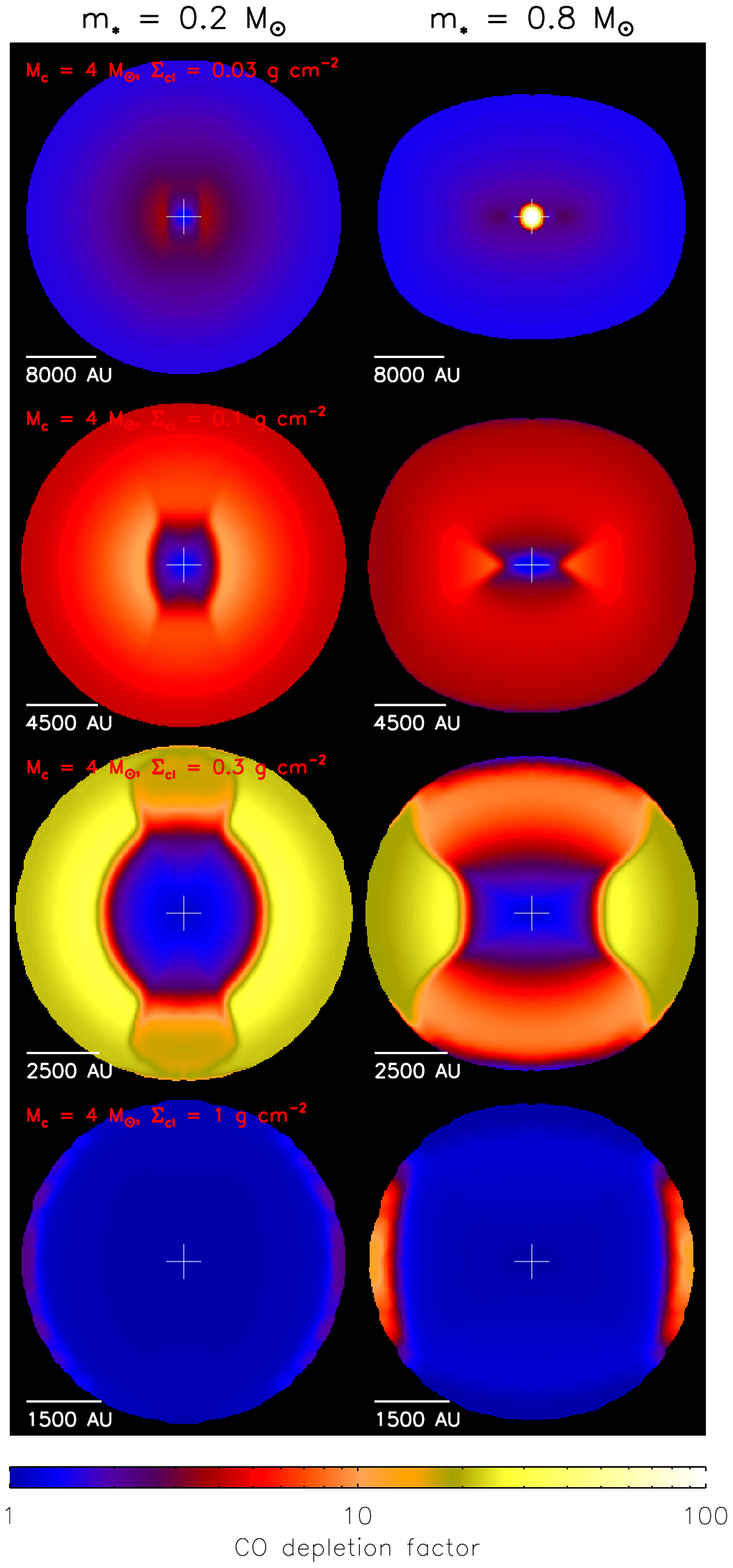}\\
\caption{CO depletion factor maps ($f_{D,{\rm{CO}}}$)
for $M_c=4\;M_\odot$ cores viewed at 
$60^\circ$ inclination
angle between line-of-sight and outflow axis with $m_*=0.2\:M_\odot$
(left) and $m_*=0.8\:M_\odot$ (right) for
$\scl=0.03,\:0.1,\:0.3,\:1\;\gcm$ (top to bottom rows; note varying
size scales).}
\label{fig:depmap}
\end{center}
\end{figure}
%========================================

In several papers (\citealt[]{ZT11}, \citealt[]{ZTM13,ZTH14},
hereafter ZT11, ZTM13, ZTH14), we have developed a RT model for
massive star formation, following evolutionary sequences of protostars
forming from massive cores.
Each evolutionary track was built from three initial conditions: clump
mass surface density $\scl$; core mass $M_c$; and core
rotational-to-graviational energy ratio $\beta_c$.
Extending this model to low-mass star formation,
low-mass cores have radius
$R_c=3.3\times10^{-2}\:\mathrm{pc}\:(M_c/2\:M_\odot)^{1/2}(\scl/0.1\:\gcm)^{-1/2}$.
Note, $\scl$ enters this formula because it sets core surface pressure
via $P\simeq0.88G\scl^2$.  In a Taurus-like region with
$\scl\sim0.03\:\gcm$, the typical radius of a $2\;M_\odot$ core (which
will form a $\sim1\:M_\odot$ star if the efficiency $\sim0.5$) is
$\sim0.06\:\rm{pc}$. In an Orion-like region with $\scl\sim0.3\;\gcm$,
this core has $R_c\sim0.02\;\rm{pc}$, consistent with observed
prestellar cores in these regions (\citealt[]{Ward07}). The core is
assumed to be a singular polytropic sphere at start of collapse, with
power-law density distribution $\rho\propto{r}^{-k_\rho}$ with
$k_\rho=1.5$.

Collapse follows inside-out expansion wave solutions
(\citealt[]{MP97}) expected of singular polytropic cores.  In later
non-homologous stages after the expansion wave reaches the core
boundary, we simply lower the envelope density while fixing the
density profile shape and outer boundary,
normalizing to remaining envelope mass. Inside the sonic point, we
apply the \citet[]{Ulrich76} solution for density and velocity profiles
to model infall with rotation, which conserves angular momentum. The
disk is described by an $\alpha$-disk model, 
including effects of outflow and accretion infall (ZTM13). 
Disk mass is set equal to 1/3 of stellar mass,
expected if self-gravity is regulating accretion and angular momentum
transport. Disk outer radius is set by the centrifugal radius,
increasing with protostellar growth as
$r_d=84\;\mathrm{AU}\;(\beta_c/0.02)(M_{*d}/m_{*d})(M_{*d}/M_c)^{2/3}(R_c/0.03\;\rm{pc})$
(ZTH14), where $m_{*d}$ is star+disk mass, and $M_{*d}$ is idealized
star+disk mass in the case of no outflow feedback, making
$m_{*d}/M_{*d}$ the average star formation efficiency.
This assumes angular momentum is conserved inside the sonic point. The
disk could be smaller
if magnetic braking is relatively strong (\citealt[]{Li14}).

Bipolar protostellar outflows carve out low-density cavities,
which gradually widen, affecting how the envelope is irradiated by the
protostar.  We include this process by considering when enough wind
momentum is accumulated to accelerate envelope material to the escape
velocity (\citealt[]{MM00}; ZTH14). We assume outflow momentum
distribution with polar angle $\dot{p}\propto1/\sin^2\theta$, suitable
for X-winds or disk-winds at large distance
(\citealt[]{MM99}). Concentration of wind momentum towards
$\theta=0^\circ$ implies outflow breakout first occurs at the poles
and then gradually widens.  Outflowing mass rate is assumed as
$0.1\;\dot{m}_*$,
typical for disk winds (\citealt[]{KP00}). After the opening angle is
determined, we carve out such cavities 
from the envelope, but with detailed wall shapes set to follow
inflowing Ulrich streamlines for simplicity. We fill the cavities with
dust and gas distributions from a disk wind solution
(\citealt[]{BP82}, ZTM13).

Protostellar evolution, i.e., internal structure, size and luminosity,
is calculated with a multi-zone numerical model (\citealt[]{Hosokawa09,
  Hosokawa10}) with accretion rate regulated by gradually widening
outflow feedback (ZTH14). 
Half the accretion energy ($Gm_*\dot{m}_*/2r_*$) is released from the
disk, partly radiated ($L_\mathrm{acc,disk}$) and partly driving the
disk-wind. The other half is radiated from the boundary layer where
the accretion flow hits the protostellar surface
($L_{\mathrm{acc,bdl}}$).  The total luminosity from the source
therefore is
$L_\mathrm{tot}=L_*+L_{\mathrm{acc,bdl}}+L_\mathrm{acc,disk}$.  

Thus we self-consistently model core collapse and rotation, disk
growth, outflow cavity widening and protostar evolution, based on
three initial conditions $\scl$, $M_c$ and $\beta_c$. We assume
typical $\beta_c=0.02$ from observations of low-mass prestellar cores
(\citealt[]{Goodman93}). $\beta_c$ affects disk size and shape of
rotating inflow streamlines, which although influencing temperatures
of inflowing gas near the disk, does not affect envelope temperature
as significantly as $\scl$. Therefore, 
we defer exploration of $\beta_c$ variation to a future study. We
choose $\scl=0.03,\:0.1,\:0.3,\:1\:\gcm$ mimicking environments from
Taurus-like regions to IRDCs/GC. We set $M_c=2,\:3,\:4\:M_\odot$ to
form a small model grid.

For each initial condition, we sample $\gtrsim100$ stages, evenly
distributed in time to construct an evolutionary sequence. The first
steps start with protostars of $m_*\sim0.01\:M_\odot$ and
$L_*\sim0.001\:L_\odot$. However, depending on $\scl$, initial
accretion luminosities vary greatly, so total stellar+accretion
luminosities range from $\sim0.1-5\;L_\odot$.  At each time step
we simulate continuum RT using Monte Carlo code
HOCHUNK3D \citep[]{Whitney13} to calculate temperature profiles.
Simulations use a 3D grid starting from the stellar
surface and covering the whole core.
The disk is typically covered by $\sim1/3-1/2$ of total radial grid of 1000 zones
(finer spacing in inner disk).
Four dust opacity types are used in envelope,
high density disk, low density disk, and outflow cavity regions
(see ZT11 for details).
Envelope and disk dust grains have ice mantles; outflow cavity grains are bare.
Gas-to-dust mass ratio of 100 is used throughout.
%ycz
Photons composing total $L_*+L_{\mathrm{acc,bdl}}$ are radiated isotropically
from the stellar surface following blackbody spectrum of temperature
$T_{*,{\mathrm{bdl}}}=[(L_*+L_{\mathrm{acc,bdl}})/4\pi\sigma{r}_*^2]^{1/4}$.
Photons composing total $L_\mathrm{acc,disk}$ are launched from the disk with 
an $\alpha$-disk radial dependence and a gaussian vertical distribution (ZTM13).
%yz

We include external illumination of a standard solar
neighborhood interstellar radiation field (ISRF)
attenuated by clump extinction of given $\scl$. This may
underestimate external illumination in some cases like the GC, where
ISRF can be $\sim1000\times$ greater 
(\citealt[]{Clark13}), though high $\scl$ and clump extinction
counteracts this. In Orion-like regions, massive young stars may
strongly increase external illumination of cores. Higher
external illumination in such high $\scl$ regions would further
amplify the contrast of the temperatures of the protostellar cores in
the low and high $\scl$ environments. Our adopted external
illumination sets a lower limit for earliest stage core temperature
$\sim10\:{\rm{K}}$.

At each time step, spectral energy distributions (SEDs) and IR
continuum images are calculated from input density and converged
temperature profiles.  
From the velocity field we can also track material parcels from
any initial core position following infall streamlines, and
calculate temperature and density evolution of that parcel ($T(t)$,
$\rho(t)$), needed for further chemical modeling.

Given CO's important role in studying molecular clouds and in
astrochemistry, we focus on modeling CO depletion as an example of the
effect of the star-forming environment on chemical evolution. We
consider CO depletion onto dust grains, thermal desorption, and cosmic
ray (CR) desorption. We adopt timescales for depletion ($\tauon$) and
thermal desorption ($\tauoffth$) from \citet[]{Visser09} (Eqs. 
35 and 36) for grains with radius $0.1\:\micron$ and abundance
$10^{-12}$ relative to H nuclei, and unity 
sticking efficiency.  For such grains, depletion timescale is
$\tauon\simeq10^4\:\mathrm{yr}\:(n_\mathrm{H}/10^5\:\mathrm{cm}^{-3})^{-1}(T/20\:\mathrm{K})^{-1/2}$,
and thermal desorption timescale is
$\tauoffth\simeq3\times10^4\;\mathrm{yr}\;[\exp(20\mathrm{K}\;/T-1)]^{55}$.
For the latter, we also assume CO binding energy onto grain ice
mantles of $E_\mathrm{CO}=1100\;\mathrm{K}$, suitable for an ice
surface of H$_2$O and CO with abundance ratio 30. The CR
desorption timescale ($\tauoffcr$) is from \citet[]{KC08} (Eq. 11;
also \citealt[]{HH93}). For the same grains and assuming CR
ionization rate $3\times10^{-17}\;\mathrm{s}^{-1}$,
$\tauoffcr\simeq2\times10^5\;\mathrm{yr}$, independent of temperature
and density.  
Note in a magnetized core/clump, CRs may be absorbed or mirrored,
lowering ionization rates by factors $\sim2-4$ (\citealt[]{PG11}), thus
increasing $\tauoffcr$ by the same factor.

Evolution of relative gas phase CO abundance ($\cogas$) is then
described via
$d\cogas/dt=-\cogas/\tauon+(1-\cogas)/\tauoffth+(1-\cogas)/\tauoffcr$.
The shortest timescale among the three processes is at most
$\sim10^4\:{\rm{yr}}$, much shorter than the star formation or
free-fall timescale, $\sim10^5\;\mathrm{yr}$ in the fastest collapse
(highest $\scl$) model. So typically we can assume the equilibrium
value for $\cogas$, calculated as
$\cogas=(\tauon\tauoffth+\tauon\tauoffcr)/(\tauon\tauoffth+\tauon\tauoffcr+\tauoffth\tauoffcr)$.
Note, $\tauoffth$ is very temperature sensitive. Below
$\sim20\;\mathrm{K}$, $\tauoffth$ is so long that it is unimportant
and CO depletion is determined by the other two processes,
$\cogas\simeq\tauon/(\tauon+\tauoffcr)$. In this case higher
temperature and density lead to shorter $\tauon$ and lower gas-phase
CO abundance (higher depletion factor), but the dependence is weak
($\tauon\sim{n}_\mathrm{H}^{-1}T^{-1/2}$).  When the temperature
reaches $\gtrsim20\;\mathrm{K}$, $\tauoffth$ drops very quickly and
becomes much shorter than the other two processes, leading to
$\cogas\simeq1$, i.e., most CO molecules in gas phase. Depletion
factor is defined $f_D\equiv1/\cogas$. Note, we always assume the same
temperature for dust and gas, valid for
$n_\mathrm{H}\gtrsim10^5\;\mathrm{cm}^{-3}$ \citep{Doty97}, 
true for envelopes in most models, except the outer envelope in the
$\scl=0.03\;\gcm$ case. Here we expect slightly higher dust than gas
temperatures because of dust absorption of IR radiation. Then,
$\tauon$ becomes longer.  Dust and gas temperature coupling is even
weaker in the outflow cavities, but CO is expected to be mostly in the
gas phase and here we anyway focus on the envelope.

\section{Results}
\label{sec:result}

Figure~\ref{fig:history} shows evolution of instantaneous star
formation efficiency ($\epsilon_*$), protostellar accretion rate, and
total luminosity in models with different $\scl$ and $M_c$.  Here
$\epsilon_*\equiv\dot{m}_*/\dot{M}_{*d}$ with $\dot{m}_*$ being
protostellar accretion rate and $\dot{M}_{*d}$ being idealized mass
growth rate of star+disk in absence of outflow feedback.  As the
protostar grows, gradual opening up of outflow cavities causes
$\epsilon_*$ to drop. Such feedback curtails the power-law growth of
$\dot{m}_*$ with $m_*$ (expected for collapse of the adopted singular
polytropic core).
This accretion rate drop marks end of main accretion phase and start
of core clearing phase (defined as $m_*=M_{\rm{env}}$).  During the
whole evolutionary track, the accretion luminosity significantly
exceeds the intrinsic internal stellar luminosity, so total luminosity
closely tracks accretion rate evolution, except for minor features due
protostellar radius evolution.  We see both accretion rate and
luminosity depend sensitively on $\scl$, but less so on $M_c$. In the
highest pressure, $\scl=1\;\gcm$ environments, accretion rate and
luminosity can be $10\times$ higher than in the $\scl=0.03\;\gcm$
environment.  Star formation timescale is also sensitive to $\scl$,
being much shorter in a high $\scl$ environment, since
$t_{*f}\propto{M}_c^{1/4}\scl^{-3/4}$ (\citealt[]{MT03}).  For a core
of given initial mass, final stellar mass also depends on $\scl$, with
a more massive star (i.e., higher star formation efficiency) formed at
higher $\scl$.  This is because it is more difficult for outflows to
break-out and widen in a denser core.

Figure~\ref{fig:trho3d} shows examples of input density profiles for
the RT simulation and resultant temperature profiles at selected
stages ($m_8=0.2,\:0.8\:M_\odot$) in two models with different
$\scl=0.03,\:1\:\gcm$, chosen to illustrate extremes of the explored
range. Although the ambient clump is important for setting core
properties, it is not included explicitly in our RT simulation (i.e.,
assumed empty outside core boundary), given the extra parameters
needed to describe clump density structure. This may lead to a modest
underestimation of core temperatures due to back heating from the
clump.  In both cases, following the evolutionary sequences, we see
that outflow cavities widen and core envelopes are being heated up due
to increasing luminosities (from 70 to $80\:L_\odot$ in low $\scl$
case; from 160 to $390\:L_\odot$ in high $\scl$ case).  At same
$m_*$, for high $\scl$ the core is much more compact and
denser than for low $\scl$.  
The outflow opening angle is also smaller in the high
$\scl$ case.  Envelope temperatures in this case reach
$\sim20$--$30\:{\rm{K}}$, significantly warmer than in the low $\scl$
case, which is $\lesssim15\:{\rm{K}}$.

Figure~\ref{fig:sedimg} shows simulated SEDs and Spitzer
$3.6,\:8,\:24\:\micron$ 3-color images at the same two evolutionary
stages for models with $M_c=4\;M_\odot$ but four different $\scl$
values. For core with given initial mass and at same $m_*$, from low
to high $\scl$ environments, the FIR SED peak becomes higher and moves
to longer wavelengths due to higher luminosity and higher envelope
extinction. Higher extinctions also suppress short wavelength fluxes,
making the MIR slope steeper than in low $\scl$ environments. IR
morphology is also strongly affected by $\scl$. For low $\scl$,
emission is dominated by the central source and the innermost regions
of the envelope and outflow cavity walls. Outflow cavities become the
dominant feature as $\scl$ increases. Due to higher extinction of the
envelope at higher $\scl$, the contrast of near- and far-facing
outflow cavities becomes higher than for lower $\scl$s. The source
also appears redder in higher $\scl$ environments. As the protostar
evolves, shorter wavelength fluxes increase while the FIR peak
decreases, which agrees with the transition from Class 0 to Class I
sources (\citealt[]{Andre95}).  Outflow cavity widening can be clearly
seen from IR morphologies. Due to larger outflow cavities and less
dense envelopes, the emission becomes more peaked towards the central
source in later stages.

Quantitative dependence of core density and temperature on $\scl$ is
shown in the first two rows of Figure~\ref{fig:evo}. Both
mass-weighted mean density and temperature are highly dependent on
$\scl$, but less affected by $M_c$. Cores in higher $\scl$
environments are denser and warmer. In the highest $\scl$ environment,
envelope mean temperature reaches $\sim30\:{\rm{K}}$, while for low
$\scl$, mean temperature is always between
$10-15\:{\rm{K}}$. Potential crowding of stars in high $\scl$
environments could make external illumination significantly higher
than the standard ISRF, which would make the envelope even warmer than
estimated here, amplifying the contrast between these
environments. Note in a high $\scl$ environment the star formation
timescale is much shorter so material stays in a warmer state for a
shorter time, while in the low $\scl$ environment it stays in a
cooler state for a longer time.  For each evolutionary track, the
envelope density decreases as a power-law as expected for a collapsing
polytropic sphere, and quickly decreases in the late stages after the
expansion wave reaches the core boundary and self-similarity breaks
down. The temperature evolution shows an increase at first and then a
drop, mainly due to increase and decrease of the accretion rate
and thus luminosity.  Outflow cavity widening also contributes to
the increase of envelope temperature, as a wider cavity facilitates
irradiation by the protostar.

Average envelope CO depletion factors, $f_{D,{\rm{CO,env}}}$, are
shown in the third row of Figure~\ref{fig:evo}, determined by balance
between depletion of molecules onto dust grains and thermal or CR
desorption.  When $T\gtrsim20\:{\rm{K}}$, thermal desorption is much
faster than the other two processes, leading to very low CO
depletion. Thus models with highest $\scl$ have lowest $f_D$.  When
$T\lesssim20\:{\rm{K}}$, thermal desorption becomes much slower and CO
depletion is determined by the other two processes. Since CR
desorption timescale is independent of temperature and density, while
depletion timescale scales as
$\tauon\simeq{n}_\mathrm{H}^{-1}T^{-1/2}$, CO is more depleted when
temperature and density are higher.  Therefore models with lowest
$\scl$, which have lower densities and temperatures, also have
relatively low $f_{D,{\rm{CO,env}}}$. The two intermediate cases with
$\scl=0.1,\:0.3\:\gcm$ have relatively high CO depletion.  In models
with lower surface densities ($\scl=0.03,\:0.1\:\gcm$), mean envelope
temperature is always $\lesssim15\:{\rm{K}}$ and $f_{D,{\rm{CO,env}}}$
does not vary much during the evolution, since depletion timescale
is only weakly dependent on temperature and density. In the two higher
surface density environments, CO depletion drops faster at first as
the temperature gradually increases to $\sim20\:{\rm{K}}$, and then
increases again as the envelope cools down. CO depletion drops at the
end of the evolutionary tracks as the density quickly declines. CO
depletion is not so sensitive to core mass.  These different behaviors
of CO depletion affect line fluxes of CO and its isotopologues, as
shown in the fourth row of Figure \ref{fig:evo}.  For cores with the
same mass, $\rm{C}^{18}\rm{O}$ line emission (assumed optically thin)
is stronger in an environment with either very high or very low
$\scl$, but lower in intermediate environments.  Variation can be
a factor $\sim5$ between cores in different environments.

Figure~\ref{fig:depmap} shows simulated CO depletion factor maps of
cores in different $\scl$ environments at the two fiducial
evolutionary stages. As discussed above, from $\scl=0.03$ to
$0.3\:\gcm$, as $\scl$ increases, $f_{D,{\rm{CO}}}$ increases because
higher core temperatures and densities in higher $\scl$ environments
shorten the depletion timescale.  However, near the protostar,
depletion is low where $T\gtrsim20\:{\rm{K}}$ cause thermal
desorption timescales to be shorter than depletion timescales.
For highest $\scl=1\;\gcm$ almost all the envelope becomes warm enough
for efficient thermal desorption 
and low $f_{D,{\rm{CO}}}$. Therefore at both extremes of high and low
$\scl$ environments, we expect smooth distributions of low CO
depletion factor over the protostellar core, but in environments with
intermediate $\scl\sim0.1$--$0.3\;\gcm$, we expect higher CO depletion
towards the outer core and a low CO depletion hole towards the
center. Such distributions of $f_{D,{\rm{CO}}}$ may be tested
observationally.

\section{Conclusions}
\label{sec:conclusion}

We investigated how mass surface densities, $\scl$, of star-forming
regions affect thermal and chemical evolution of protostellar cores,
with implications for infrared morphologies and CO depletion. In high
$\scl$, high-pressure environments, cores of given mass are more
compact, denser, collapse with higher accretion rates, forming more
luminous protostars with higher efficiencies that lead to warmer
infall envelopes.  When $\scl=0.03\;\gcm$, the mean temperature is
$\sim10-15\:{\rm{K}}$. It rises to $\sim25-30\:{\rm{K}}$ in the
highest $\scl=1\;\gcm$ case.  As $\scl$ increases, infrared
morphologies reveal more dominant outflow cavities that show greater
asymmetry between near- and far-facing sides.  CO depletion factors
are small at low $\scl$, since cores have low densities leading to low
rates of adhesion of molecules to dust. $f_{D,{\rm{CO}}}$ rises at
intermediate $\scl$ as cores become denser, but there is a central
warm, $\gtrsim20\:{\rm{K}}$ low-depletion region. This grows for more
luminous protostars and for the high $\scl\sim1\;\gcm$ case it
encompasses the entire core envelope over most of the evolution.

Further implications of these models for: disk structure and possible
fragmentation for binary formation (\citealt[]{Kratter10}) with
fragmentation via gravitational instability being easier in larger,
cooler disks; more general astrochemical evolution
(\citealt[]{Visser11}) in the infall envelope and disk; and for dust
grain coagulation in these phases (\citealt[]{Birnstiel12}), remain to
be investigated in future studies.

\acknowledgements We thank Paola Caselli for discussions.

\end{document}